\documentclass[aps,prx,twocolumn,nofootinbib]{revtex4-1}
\usepackage[english]{babel}
\usepackage[utf8]{inputenc}
\usepackage[pdftex]{graphicx}
\usepackage{hyperref}
\usepackage{url}
\usepackage[titletoc,title]{appendix}
\usepackage[suffix=]{epstopdf}
\usepackage{amsmath}
\usepackage{amssymb}
\usepackage{dsfont}
\usepackage{mathtools}
\usepackage{bbm}
\usepackage{xcolor}

\def\be{\begin{equation}}
\def\ee{\end{equation}}
\def\bea{\begin{eqnarray}}
\def\eea{\end{eqnarray}}

\def\bra#1{\mathinner{\langle{#1}|}}
\def\ket#1{\mathinner{|{#1}\rangle}}

\def\ul#1{{\underline{#1}}}

\newcommand{\ave}[1]{{\left\langle #1\right\rangle}}

\newcommand{\ii}{ {\rm i} }

\newcommand{\ZZ}{\mathbb{Z}}

\def\tr{{{\rm tr\,}}}

\let\oldref\ref
\renewcommand{\ref}[1]{(\oldref{#1})}

\begin{document}
	
	\title{
	Many-body quantum chaos: Analytic connection to random matrix theory}
	
	\author{Pavel Kos}
	\affiliation{Physics Department, Faculty of Mathematics and Physics, University of Ljubljana, Jadranska 19, SI-1000 Ljubljana, Slovenia}	
	\author{Marko Ljubotina}
	\affiliation{Physics Department, Faculty of Mathematics and Physics, University of Ljubljana, Jadranska 19, SI-1000 Ljubljana, Slovenia}
	\author{Toma\v z Prosen}
	\email{Corresponding author. email: tomaz.prosen@fmf.uni-lj.si}
	\affiliation{Physics Department, Faculty of Mathematics and Physics, University of Ljubljana, Jadranska 19, SI-1000 Ljubljana, Slovenia}

\begin{abstract} 
A key goal of quantum chaos is to establish a relationship between widely observed universal spectral fluctuations of clean quantum systems and random matrix theory (RMT). 
Most prominent features of such RMT behavior with respect to a random spectrum, both encompassed in spectral pair correlation function, are statistical suppression of small level spacings (correlation hole) and 
enhanced stiffness of the spectrum at large spectral ranges.
 For single particle systems with fully chaotic classical counterparts, the problem has been partly solved by Berry [Proc. R. Soc. London, A400, 229 (1985)] within the so-called diagonal approximation of semiclassical periodic-orbit sums,
while the derivation of the full RMT spectral form factor $K(t)$ (Fourier transform of spectral pair correlation function), from semiclassics has been completed by 
M\" uller et al. [Phys. Rev. Lett. 93, 014103 (2004)]. In recent years, the questions of long-time dynamics at high energies, for which the full many-body energy spectrum becomes relevant, are coming at the forefront even for simple many-body quantum systems, such as locally interacting spin chains. Such systems display two universal types of behaviour which are termed as the `many-body localized phase' and `ergodic phase'. In the ergodic phase, the spectral fluctuations are excellently described by RMT,  
even for very simple interactions and in the absence of any external source of disorder. Here we provide a {clear} theoretical explanation for these observations.  
We compute $K(t)$ in the leading two orders in $t$ and show its agreement with RMT 
for non-integrable, time-reversal invariant many-body systems without classical counterparts, a generic example of which are Ising spin 1/2 models in a periodically kicking transverse field.
In particular, we relate $K(t)$ to partition functions of a class of twisted classical Ising models on a ring of size $t$,
hence the leading order RMT behavior $K(t) \simeq 2t$ is a consequence of translation and reflection symmetry of the Ising partition function.
\end{abstract}

	\maketitle

\section{Introduction}

RMT was introduced into physics in the 1950s by Wigner \cite{Wigner} for providing a statistical description of nuclear resonance/excitation spectra. It should be intuitively clear that a system consisting of a few tens of nucleons coupled via short and long-range interactions is complicated enough that a successful description of experimental spectral fluctuations in terms of an ensemble of random Hamiltonians with independent stochastic matrix elements is not that surprising. An example of a robust phenomenological measure of fluctuations is the statistical variance of the number of energy levels in an interval of fixed length $\Delta E$ which, in RMT and experimental nuclear spectra \cite{Haq}, grows as $\sim \log |\Delta E/\bar{\rho}|$ 
({known as spectral stiffness}), rather than $\sim\sqrt{\Delta E/\bar{\rho}}$ as in the Poissonian random spectrum ($\bar{\rho}$ is the average density of states). The atomic spectra observed already by 1960 exhibited the so-called `level repulsion' which can be quantitatively explained \cite{Rosenzweig} with Wigner's RMT. However, in the early 1980s a much more surprising fact has been revealed, namely that RMT also works extremely well for capturing spectral fluctuations of simple single-particle systems whose corresponding classical dynamics are completely chaotic, such as dispersive (Sinai) billiards or hydrogen/Rydberg atoms in external magnetic or microwave fields. These observations {\cite{MacDonald,CGV80,Berry81}, 
termed as the {\em quantum chaos conjecture} which has been concisely stated in \cite{BGS84}}, have driven the field of quantum chaos for decades.
The first, partial explanation for the success of RMT in simple chaotic systems came from Berry's semiclassical (small effective $\hbar$) calculation \cite{Berry85} of the spectral form factor $K(t)$ {in terms of a double sum over classical unstable periodic orbits, which we shall explain below.
$K(t)$ is defined as a Fourier transformation of the two-point correlation function of the spectral density {$\rho(E)=\sum_{j}\delta(E-E_j)$, $\{ E_j\}$ being the energy spectrum}
\be
K(t) = \int e^{-\ii \epsilon t/\hbar} \left(\ave{\rho(E+\frac{\epsilon}{2})\rho(E-\frac{\epsilon}{2})}_{\!E}\!-\,\bar{\rho}^2\right) \frac{{\rm d}\epsilon}{\bar{\rho}}, \label{eq:SFF}
\ee
where $\bar{\rho} =\ave{\rho(E)}_E$ {and $\ave{\ldots}_E$ represents local energy average over an energy shell (say of width $\Delta E$) containing many levels $\bar{\rho}\, \Delta E \gg 1$, in case of autonomous (time-independent) systems.
In case of periodically driven, i.e. Floquet systems that we shall discuss later in this paper, the average over the full range $[0,2\pi)$ of quasi-energies -- eigenvalues of unitary Floquet (one-period) propagator $U$
is normally considered as physical properties are not expected to depend on the particular value of quasi-energy.}
A fruitful intuition stems from observation that $K(t)$ characterizes all pair-correlation 
properties including the level repulsion and spectral stiffness, since in RMT $\{E_j\}$ can be considered as a 
fictitious one dimensional (Dyson's) gas with a logarithmic pairwise interactions \cite{Haake}.

For integrable systems, possessing a complete set of conserved quantities, the energy spectrum $\{E_j\}$ 
is conjectured \cite{BerryTabor} to represent a Poisson random uncorrelated sequence, so the spectral form factor (\ref{eq:SFF})
can be exactly computed as $K(t)\equiv t_{\rm H}{=2\pi\hbar\bar{\rho}}={\rm const}$ for all $t>0$, and thus provides a clear discriminator between integrable and chaotic systems, since for the latter 
$K(t)\propto t$, {in agreement with explicit predictions of RMT}, as we shall explain in Subsect.~\ref{semiclassK} below.

A clear heuristic derivation of RMT spectral form factor $K(t)$ for classically strongly chaotic (hyperbolic) systems from semiclassical periodic orbit theory, starting from Berry's diagonal approximation \cite{Berry85}, 
upgraded to second order in $t$ by Sieber and Richter \cite{SR2001,S2002}, and finally completed to all orders in a tour de force by M\" uller et al. \cite{Haake2004,Haake2005}, has been arguably the main
accomplishment of the field of quantum chaos of single or few particle systems.
Nevertheless, a rigorous proof of the quantum chaos conjecture has so far only been possible for a much more abstract class of single-particle systems, specifically for mixing quantum graphs \cite{Weidenmuller1,Weidenmuller2}.
These semiclassical periodic-orbit approaches have a natural generalisation to a quantum many-body problem 
for bosons when the number of quanta per mode is large \cite{Engl,Urbina,Guhr,Remy}.

However, RMT has also been found to excellently describe spectral fluctuations in the simplest, say low-dimensional and locally interacting, non-integrable many-body systems where local degrees of freedom have no classical limit at all, such as spins $1/2$, qubits, fermions etc. \cite{ergodic1,ergodic2,ergodic3,ergodic4} and where no semiclassical or mean-field approach can be applied.
{Due to such phenomenological success, RMT statistics of level spacings is nowadays used essentially as a definition of the so-called quantum chaotic or ergodic phase (see e.g. \cite{Huse,Serbyn,Luitz,Papic,Sondhi,Bloch}). Moreover, the ergodic phase has been intensively theoretically investigated in recent years and its most concise characterization
is provided by the so-called eigenstate thermalization hypothesis (see e.g., \cite{ETH,Huse}).
Nevertheless,} there has so far been no {proposition} of the underlying dynamical (or microscopic) mechanism (such as unstable periodic orbit pairings in semiclassical chaotic systems discussed above).
Recent studies of out-of-time-ordered correlations in many-body systems, some of which establish exponential growth (in particular in 0+1 
dimensional systems such as the Sachdev-Ye-Kitaev model), have no clear connection to Lyapunov instability as it is understood in classical dynamical systems theory, and is the only mathematically meaningful definition of chaos. This has to do with the lack of the concept of classical orbits and the corresponding unstable (non-linear) equations of motion which result in sensitive dependence on initial conditions (e.g. the butterfly effect). In short, the concept of orbits and Lyapunov chaos does not make sense at `$\hbar\sim 1$'. One thus
urgently needs alternative concepts which would enable one to explain the surprising success of RMT in simple many-qubit systems.

{Providing one such concept is the main objective of this article.}
 We identify a coherent structure, in a class of generic many-body quantum systems {with the lowest, 
two-dimensional local Hilbert space (qubits, or spins 1/2)}, which is responsible for building up level (spectral) correlations. 
Expanding $K(t)$, which is written 
as the product of two traces of the quantum mechanical propagator (see Subsect.\ref{semiclassK}), in the computational spin basis and writing it in a discrete-path-integral like fashion, we find that the 
leading contribution comes from constructive interference which corresponds to a partition function of a classical one-dimensional Ising model. Furthermore, sub-leading contributions 
can be interpreted as a family of partition functions of so-called twisted Ising models which are classified using a novel diagrammatic technique. In terms of this expansion the leading contributions to $K(t)$ are shown to exactly correspond to RMT results {$K(t) \simeq 2t$} for times longer than a certain crossover time $t^*$, while the non-universal results at $t < t^*$ are shown to reproduce  numerical data
extremely well. The time-scale $t^*$, which scales logarithmically with the system size, can be interpreted as a quantum many-body analogue of the Ehrenfest ({or Thouless}) time. {
Finally, we identify the non-semiclassical analog of the Sieber-Richter pairing mechanism \cite{SR2001} and exactly reproduce the 
sub-leading RMT term $-2t^2/t_{\rm H}$ as well.}

\subsection{Spectral form factor in Floquet systems and periodic orbit theory}

\label{semiclassK}

In order to address the setup with a minimal amount of inessential technical complications we decide to study periodically driven (Floquet) many-body systems in the absence of
any conserved charges or unitary symmetries. Even in single-particle context these are the minimal models of quantum chaos \cite{Haake}
and correspond to Dyson's circular ensembles of random unitary matrices \cite{Mehta}. 

In this subsection we define the main object of our study, namely the spectral form factor for Floquet systems, and for comparison with the main derivation in Sect.~\ref{main} of our paper, outline the key steps of historical semiclassical derivation of the RMT form factor in terms of periodic orbit theory. 
For a unitary one-period Floquet propagator $U$ we write the eigenphases $\varphi_n$ and eigenvectors $\ket{n}$ as $U\ket{n}=e^{-\ii\varphi_n}\ket{n}$, $n=1,\ldots,{\cal N}$, {where ${\cal N}$ denotes
the dimension of the Hilbert space.}
	The spectral density (1-point function) is now defined as
	\be
	\rho(\varphi)=\frac{2\pi}{\cal N}\sum_n \delta(\varphi-\varphi_n),
	\ee
	and is normalized to a unit mean level density 
	\be
	\ave{\rho(\varphi)}_\varphi \equiv \frac{1}{2\pi} \int_0^{2\pi}\!{\rm d}\varphi\,\rho(\varphi)=1.
	\ee
	Locally averaged density is expected to be} $\varphi$-independent which makes
	Floquet systems particularly appealing for studying spectral fluctuations. These are encapsulated in the
        connected (2-point) spectral correlation function 
	\be
	R(\vartheta) = \ave{\rho(\varphi+\vartheta/2)\rho(\varphi-\vartheta/2)}_\varphi-\ave{\rho}_\varphi^2
	\ee
	which is, again, expected to be homogeneous ($\varphi$-independent).
	An equivalent, and very convenient quantity is the spectral form factor $K(t)$, $t\in\ZZ$, defined as an appropriately scaled Fourier transform
	\bea
	&&K(t) = \frac{{\cal N}^2}{2\pi} \int_0^{2\pi}\!{\rm d}\vartheta\,R(\vartheta)e^{-\ii\vartheta t} \nonumber \\
	&&=\sum_{n}e^{-\ii t\varphi_n} \sum_{n'}e^{\ii t\varphi_{n'}} - {\cal N}^2 \delta_{t,0} = 
	\left| \tr U^t\right|^2 - {\cal N}^2\delta_{t,0}.\;\;  \label{eq:FF}
	\eea
	Finally, one writes 
	\be
K(t) = \ave{ (\tr U^t) (\tr U^{-t})} - {\cal N}^2 \delta_{t,0},
\label{K}
\ee
where  $\ave{...}$ represents an appropriate additional averaging, either over local windows of time $t$ (moving time average) or over an ensemble of similar systems, which is needed since the spectral form factor (\ref{eq:FF}) is not a self-averaging quantity \cite{Prange}.

For circular random matrix ensembles \cite{Mehta} which are expected to model Floquet systems in RMT
	(orthogonal/unitary ensemble (OE/UE) for systems with/without time-reversal or more general anti-unitary symmetry) the spectral form factor up to Heisenberg time, $t < {\cal N}$, reads
	\bea
	K_{\rm OE}(t) &=& 2t - t \ln(1 + 2 t/{\cal N}) = 2t - 2t^2/{\cal N} + \cdots,\quad  \label{RMTSFF}\\	
	K_{\rm UE}(t) &=& t.
	\eea
{Note that exactly the same expressions hold as well for Gaussian ensembles of RMT which model time-independent systems.}

For Floquet systems with a well defined classical limit, where the motion is hyperbolic (chaotic) everywhere in the phase-space, one can write $\tr U^t$ in terms of a Feynman path integral and evaluate it by the method of stationary phase in terms of a finite sum over all periodic orbits $p$ of length $t$, with classical actions $S_p$ and 
amplitudes $A_p$ which are proportional to inverse square root of stability exponents
\be
\tr U^t \simeq \sum_p A_p e^{-\ii S_p/\hbar}.
\ee
Note that we chose to work at fixed $t$ rather than at fixed energy $E$ as is customary in semiclassical analysis of time-independent systems.
Here one assumes that the effective $\hbar$ is small, i.e. $S_p \gg \hbar$ for all $p$, which is justified for large Hilbert space dimensions ${\cal N}\gg 1$.
A semiclassical representation of the spectral form factor can then be written as
\be
K(t) \simeq \ave{\sum_{p,p'} A_p A^*_{p'} e^{-\ii (S_p-S_{p'})/\hbar}}.
\label{KSC}
\ee
Berry identified the leading RMT contribution {$K(t) \simeq t$} from the diagonal terms of paired orbits, arguing that the non-diagonal terms of unequal orbit pairs average out in the leading order due to random phases.
{The diagonal contribution then results from Hannay-Ozorido de Almeida sum rule \cite{OdAH} $\sum_p |A_p|^2 = t$, which is just a restatement of classical ergodicity.}
For systems with {time-reversal invariance (TRI)}, the leading order of RMT result (\ref{RMTSFF}), $K(t) \simeq 2t$, then simply follows by pairing each orbit with itself {$p'=p$} and its time-reversed partner {$p'=\bar{p}$, noting that $S_{\bar{p}}=S_p$ and $A_{\bar{p}}=A_p$.} However, Berry's result only holds on time-scales much shorter than the Heisenberg time $t \ll t_{\rm H} $ which translates to spectral correlations on quasi-energy ranges much larger than the mean level spacing. That result can thus be considered as the leading order of a power series expansion (in $t$) of the RMT expression for $K(t)$.
Further progress came only 16 years later
when  Sieber and Richter \cite{SR2001,S2002} correctly identified the next-to-leading RMT term of $K(t) = 2t - 2t^2/t_{\rm H}+{\cal O}(t^3/t_{\rm H}^2)$ {for TRI systems} via the {\em self-encountering} periodic orbit doublets. {Specifically, they decomposed the periodic orbit sum into two parts, the first containing a majority of orbits which never come close to themselves before the full period and in the second part, they considered orbits which experience a close self-encounter. They argued that the orbits from the second group form doublets with very similar actions $S_p$ and amplitudes $A_p$ 
which thus coherently interfere in the double sum (\ref{KSC}) and result, after careful bookkeeping, exactly in the second order term $-2t^2/t_{\rm H}$ of RMT.}
 It took another few years of efforts until this endeavour has finally been completed in \cite{Haake2004,Haake2005} (see also \cite{Saito} for the analysis of unitary-to-orthogonal ensemble crossover) by correctly identifying all the terms in the power-series expansion of $K(t)$ from sums over chaotic periodic orbits {with an arbitrary number of self-encounters}.

\section{Partition function expansion of the spectral form factor}

\label{main}

Here, however, we consider an interacting many-body system of quantum excitations without any meaningful classical limit, so the semiclassical periodic orbit theory is not applicable.
We consider a system of $\ell$ spins $1/2$ (qubits) described by Pauli spin operators $\sigma^{(\alpha)}_{x}$, 
$\alpha\in\{1,2,3\}$, $x\in\{1,\ldots,\ell\}$, where the time-evolution is given by the following two-step unitary Floquet propagator of a periodically pulse-driven Hamiltonian 
\be
H(t) = H_{0} + H_{1} \sum_{m\in\mathbb{Z}} \delta(t-m)
\label{eq:kicked}
\ee
(time is measured in units of pulse period)
\begin{eqnarray}
U &=& {\cal T}\!\text{-}\!\exp\left(-\ii \int_0^1\!{\rm d}t\,H(t)\right) = V W,\label{prop}\\ 
W &=& e^{-\ii H_0},\; H_0 = \sum_{x} J^{1}_{x} \sigma^{(3)}_{x} + \sum_{x<x'} J^2_{x,x'} \sigma^{(3)}_{x} \sigma^{(3)}_{x'} + \cdots, \nonumber \\
V &=& e^{-\ii H_1} = v^{\otimes \ell},\qquad
H_1 = h \sum_{x} \sigma^{(1)}_{x}, \nonumber
\end{eqnarray}
where $v$ is a $2\times 2$ matrix with elements $v_{00}=v_{11}=\cos h$, $v_{01}=v_{10}=\ii\sin h$.
In the basis of ${\cal N}=2^\ell$ joint eigenstates of $\sigma^{(3)}_{x}$, $\sigma^{(3)}_{x}\ket{\ul{s}}= (-1)^{s_{x}}\ket{\ul{s}}$, labelled by classical spin configurations $\ul{s}=(s_1,\ldots,s_\ell)$, $s_{x}\in\{0,1\}$, $W$ acts as a pure phase factor
\begin{eqnarray}
W\ket{\ul{s}} &=& e^{-\ii\theta_{\ul{s}}} \ket{\ul{s}},\\ 
\theta_{\ul{s}} &=& \sum_{x} J^1_{x} (-1)^{s_{x}} + \sum_{x<x'} J^2_{x,x'} (-1)^{s_{x}+s_{x'}}+\cdots
\end{eqnarray} 
while the matrix elements of $V$ factorize 
\begin{equation}
\bra{\ul{s}}V\ket{\ul{s}'}=\prod_{x=1}^\ell
v_{s_{x},s'_{x}}.
\end{equation} 
The propagator (\ref{prop}) defines a generic family of Ising models periodically kicked with a uniform transverse field (generalizing the {\em kicked Ising chain} \cite{KI_PRE,KI_JPA}, where RMT spectral fluctuations have been verified to a high precision \cite{Pineda}). {See also Ref.~\cite{Guhr2} for a related discussion of
transfer-matrix evaluation of the many-body propagator.} 
In more abstract terms, one can also view $H_0$ as a generic integrable or many-body localized system with l-bits $\sigma^{(3)}_{x}$ and $V$ as a global perturbation. The model is time-reversal invariant as the matrices of $H_{0,1}$ are 
 {\em real}, i.e., $V,W$ are symmetric. 
 
 {We note immediately that the method that shall be developed below can be used as well to study 
 a continuous-time version of the transverse field Ising model, where the kicked model (\ref{eq:kicked}) represents its trotterization (via the Trotter formula) by substituting
 $J^k_{x...} \to (\Delta t) J^k_{x...}$, $h \to (\Delta t) h$ and carefully performing double scaling $\ell \to \infty$ and $\Delta t \to 0$ where the thermodynamic limit should be considered first.
Further, more general forms of off-diagonal perturbations $V$ can be considered by allowing an arbitrary spatial dependence of the magnetic field $h\to h_x$. Nonetheless, the particular system that we choose to study in the present paper represents a minimal generic model of many-body quantum chaos at $\hbar\sim 1$.}


\begin{figure}
\hspace{-3mm}\includegraphics[width=0.49\textwidth]{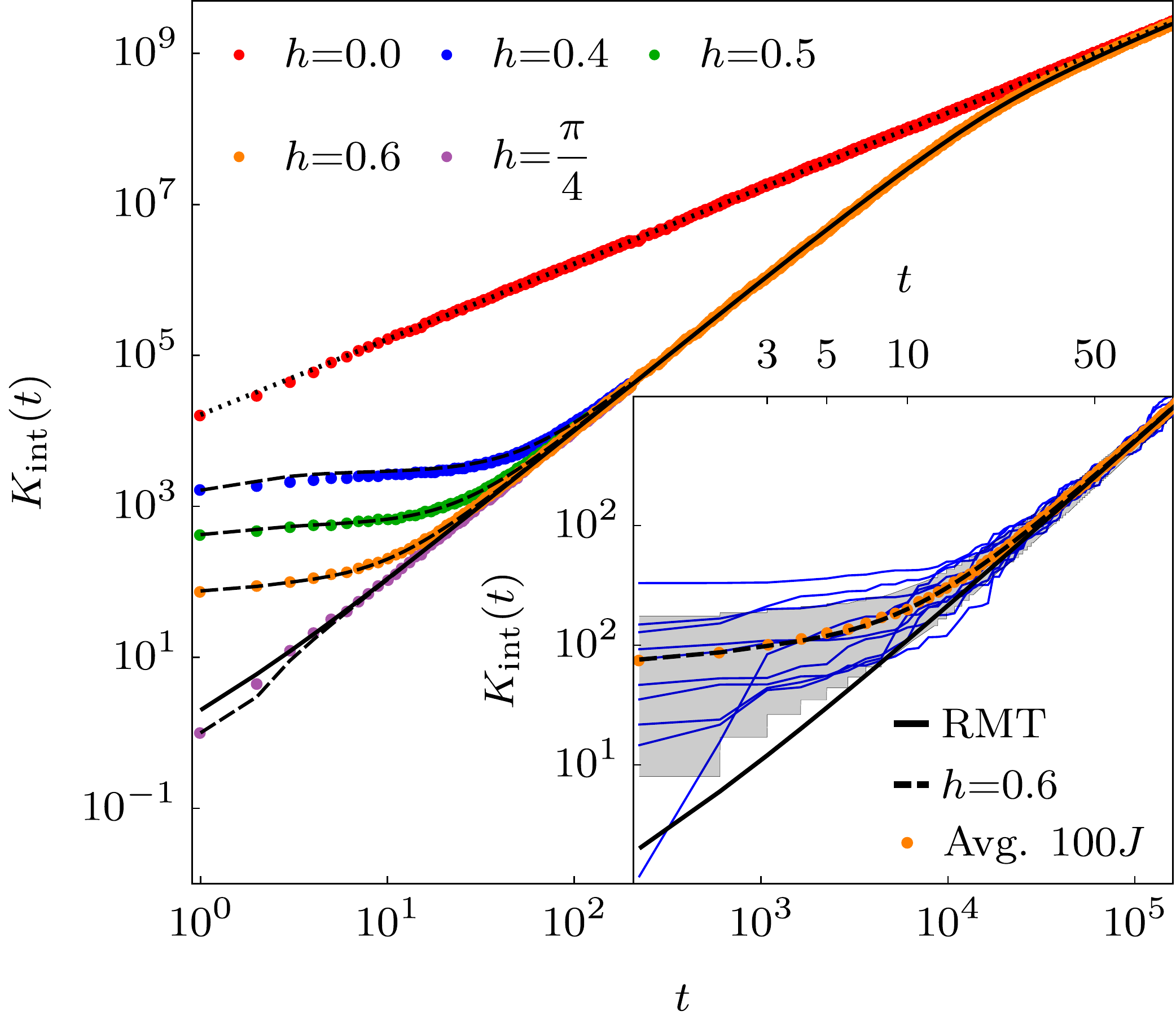}
\caption{Time integrated spectral form factor for the kicked Ising model for up to ten Heisenberg times and at four different values of the transverse field $h$. 
The black dashed lines show predictions of the random phase model, while dotted/solid lines give the Poissonian/RMT result for OE. The coloured dots show numerical data averaged over $100$ realizations of $J$ sampled uniformly in the interval $[5.5,55]$, 
for the kicked Ising model introduced in eq. (\ref{eq:LR_Ising}). 
In the inset we show that averaging is only necessary for very short times. Thin blue lines are particular realizations of the model for $J=10,11,\ldots 19$ and the shaded area shows the second to ninth decile assuming exponential distribution for realizations of $K(t)$ resulting in a hypoexponential distribution for the integrated spectral form factor.
Other parameters are fixed to $\ell=14$, $a=1$, $b=5$, $\alpha=1.5$.
}
\label{fig:K2int}
\end{figure}

\begin{figure*}
\includegraphics[width=\textwidth]{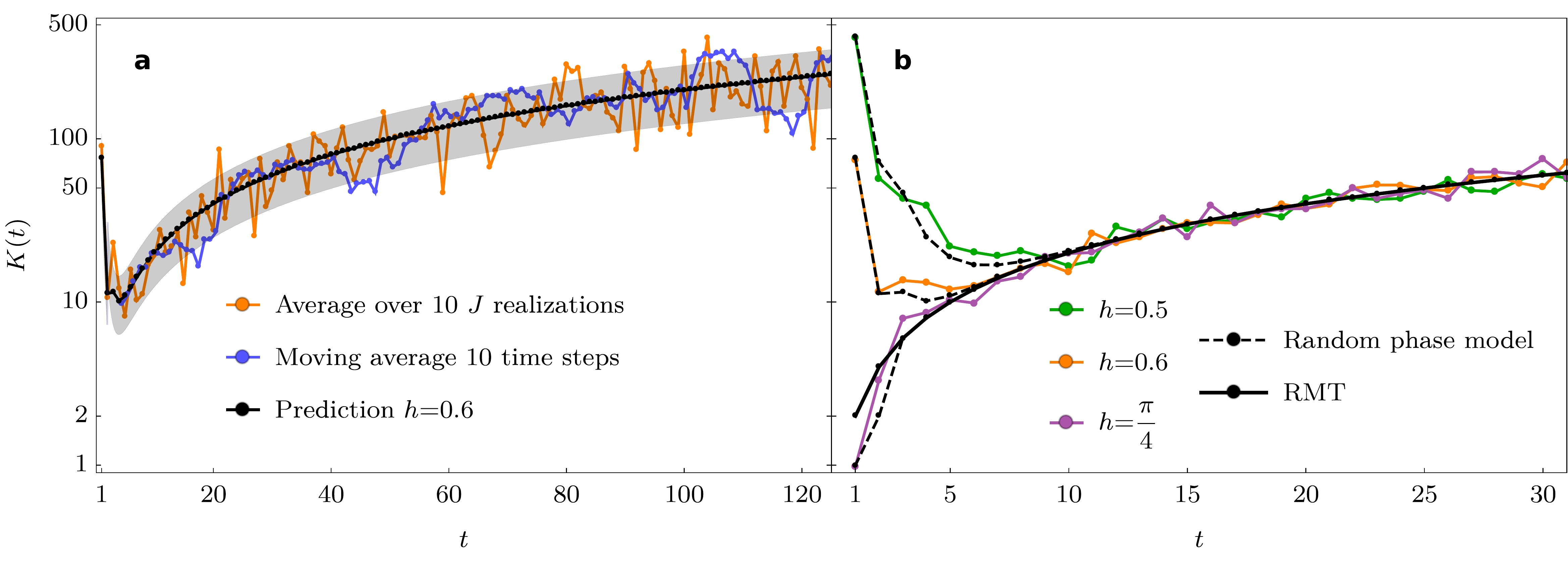}
\caption{Spectral form factor for the kicked Ising model at short times. Panel (a): Comparison of a moving average over ten consecutive time steps and fixed model realisation ($J=10$) 
with the averaging over ten realizations $J=10,11,\ldots 19$. All other parameters are fixed to $\ell=14$, $a=1$, $b=5$, $\alpha=1.5$. 
Shaded area shows the second to ninth decile assuming exponential distribution for realizations of $K(t)$ resulting in the gamma distribution for the averages.
Panel (b): $J$-averaged spectral form factor (now averaged over a sample of 100 values of $J\in[5.5,55]$) at short times shows deviations from the RMT and the deviations are well captured by the random phase model. 
}
\label{fig:K2}
\end{figure*}

We start by considering an expression (\ref{K}) for the spectral form factor of Floquet systems $K(t) = \ave{(\tr U^t)(\tr U^{-t})}$, defined for positive integer time $t$.
Inserting multiple identities $\sum_{\ul{s}_\tau}\ket{\ul{s}_\tau}\bra{\ul{s}_\tau}=1$ in $\tr U^t$ and $\sum_{\ul{s}'_\tau}\ket{\ul{s}'_\tau}\bra{\ul{s}'_\tau}=1$ in $\tr U^{-t}$, we obtain
\bea
K(t) &=& \sum_{\ul{s}_1,\ldots,\ul{s}_t} \sum_{\ul{s}'_1,\ldots,\ul{s}'_t}
\ave{e^{-\ii \sum_{\tau=1}^t (\theta_{\ul{s}_\tau}-\theta_{\ul{s}'_\tau})}} \nonumber \\
&&\times \prod_{x=1}^\ell  \prod_{\tau=1}^t v_{s_{x,\tau},s_{x,\tau+1}}v^*_{s'_{x,\tau},s'_{x,\tau+1}}.                                                                                                                                       
\eea
Note that taking the trace implies periodic boundary conditions in time $t+1\equiv 1$.
Assuming pseudo-randomness of the phases $\theta_{\ul{s}}$ one has
\be
\ave{e^{-\ii \sum_{\tau=1}^t (\theta_{\ul{s}_\tau}-\theta_{\ul{s}'_\tau})}} = \delta_{\langle \ul{s}_1,\ldots,\ul{s}_t\rangle,
\langle \ul{s}'_1,\ldots,\ul{s}'_t\rangle} + {\text{fluctuations}}
\label{eq:delta}
\ee
where $\langle \ul{s}_1,\ul{s}_2,\ldots\ul{s}_t\rangle$ represents a lexicographically ordered string of words 
$\ul{s}_1,\ul{s}_2,\ldots,\ul{s}_t$. In the ideal case, where all $2^\ell$ phases $\theta_{\ul{s}}$ can be assumed to be independent
random and uniform in $[0,2\pi)$ (which is equivalent to the assumption that all the coupling constants $J^{k}_{x,x'\ldots}$ are i.i.d.), the fluctuation term in (\ref{eq:delta}) exactly vanishes.

We shall refer to such an ideal situation in which fluctuations in (\ref{eq:delta}) are set to zero as a {\em random phase model} (RPM). 
Below we show how to compute $K(t)$ for the RPM and demonstrate that the result describes both the universal (RMT) and non-universal (short time) regimes of 
large families of clean kicked Ising models excellently. 

For times much shorter than the Heisenberg time $t_{\rm H}=2^\ell$ one may assume that all configurations 
$\ul{s}_\tau$ in the string $\ul{s}_1,\ul{s}_2,\ldots,\ul{s}_t$ are different. Then, eq. (\ref{eq:delta}) implies that there exists a permutation $\pi \in S_t : \tau \to \pi(\tau)$, such that 
$\ul{s}'_\tau= \ul{s}_{\pi(\tau)}$, therefore
\bea
K(t) &=& \sum_{\pi\in S_t} Z^\ell_\pi,\quad\text{where} 
\label{eq:Sum_of_permutations}
\\
Z_\pi &=& \sum_{s_1,\ldots,s_t} \prod_{\tau=1}^t v_{s_\tau,s_{\tau+1}} v^*_{s_{\pi(\tau)},s_{\pi(\tau+1)}},
\eea
up to ${\cal O}(t/2^\ell)$. Denoting by $w(\ul{s})=\frac{1}{2}\sum_{\tau=1}^t (1-\delta_{s_\tau,s_{\tau+1}})$ a half-number of domain-walls in a periodic spin sequence $\ul{s}$ (which is always an integer), $Z_\pi$ can be rewritten as
\be
Z_\pi = (\cos h)^{2t} \sum_{\ul{s}\in\{0,1\}^t} (-|\tan h|)^{w(\ul{s})+w(\pi(\ul{s}))}.
\label{eq:Z_pi}
\ee
Note that for the identity permutation, $Z_{\text{Id}}$ is a partition function of a classical one-dimensional Ising model (on a ring of circumference $t$) which can be calculated via a $2\times 2$ transfer matrix 
$T_{ss'}=|v_{ss'}|^2$, namely $Z_{\text{Id}}=\tr T^t = 1 + (\cos 2h)^{t}$.
$Z_\pi$ equals $Z_{\text{Id}}$ for any other permutation which does not change any neighbours in the string $\ul{s}$,
i.e. conserves the wall counting function $w(\ul{s})$. These are exactly the $t$ cyclic permutations and $t$ anti-cyclic permutations -- compositions of cyclic permutations with inversion $\tau \to t+1-\tau$.
For all other permutations $\pi$ which contain at least one pair of neighbour-changes, one can show that the {\em twisted partition functions} $Z_{\pi\neq\text{Id}}$ are strictly smaller, and can be systematically computed using a diagrammatic technique (see Appendix A). Upon approaching the thermodynamic limit $\ell \to\infty$ at fixed $t$ one thus finds an exact asymptotic result
\bea
K(t) &\simeq& 2t (1+(\cos 2h)^{t})^\ell \nonumber \\ &\simeq& 2t \quad \text{for}\quad t \gg t^*,\label{KRPM} \\ 
t^* &=& -\frac{\ln\ell}{\ln\cos 2h}. \label{tstar}
\eea
This result can be interpreted as an analogue of Berry's diagonal approximation and yields the first order of RMT [Eq.~(\ref{RMTSFF})], while exact non-universal behaviour is predicted for times $t \lesssim t^*={\cal O}(h^{-2}\ln\ell)$. The time scale $t^*$ which separates the universal from non-universal behaviour can be interpreted as a kind of quantum many-body Ehrenfest time. 
Note that in the limit $h\to 0$, {the spectral form factor grows to the saturation value ${\cal N}$ at $t\sim 1$, and} one obtains the expected Poissonian behaviour
\be
K_{h=0}(t)={\cal N},
\ee 
which is typical for completely integrable 
systems \cite{BerryTabor}. 
{
At the two points, $h=0$ and $h=\frac{\pi}{2}$, where the crossover time $t^*$ [Eq. (\ref{tstar})] formally diverges (and close to them, for finite $\ell$) our expansions in 
$t$ brakes down (see appendix A 
 and Fig.~\ref{fig3}).
The case $h=\pi/2$ actually corresponds to a generic realization of a Floquet time crystal \cite{Sondhi,time-crystal} which is non-ergodic and where the discrete translational 
invariance in time is spontaneously broken. This corresponds to a staggered behaviour of spectral form factor 
\be
K_{h=\frac{\pi}{2}}(t) = 
\begin{cases}
{\cal N}, & t \text{ even}, \\
0, & t \text{ odd}.
\end{cases}
\ee
We note that in the range $h \in \left(\frac{\pi}{4},\frac{3\pi}{4}\right)$ the RPM spectral form factor (\ref{KRPM}) in the non-RMT regime, $t<t^*$, still displays characteristic period-2
oscillations.
}

Carefully subtracting double counted terms where exactly one configuration (word) in the string (`orbit') $\ul{s}_1,\ul{s}_2,\ldots,\ul{s}_t$
appears twice one obtains exactly the RMT result Eq.(\ref{RMTSFF}) up to the second order 
\be
K(t) = 2t - 2t^2/2^\ell + {\cal O}(t^3/4^\ell)
\ee 
(see Appendix B). This can be considered as a many-qubit analogy of the Sieber-Richter pairing mechanism \cite{SR2001,S2002}. We conjecture that it should be possible
to obtain RMT result to all orders by implementing the multiple-counting technique generalizing the diagrammatics sketched in Appendixes A,B.

\section{Kicked transverse field Ising model: Theory explains numerics}
\label{Numerics}

We compare analytic results for the RPM to exact numerical computations of the spectral form factor in the following family of kicked Ising models
\be
J^1_{x} = a + \frac{N_1 b}{x^\alpha},\quad J^2_{x,x'} = \frac{N_2 J}{(x'-x)^\alpha}, \quad J^{k>2}_{x,x'\ldots}\equiv 0,\quad
\label{eq:LR_Ising}
\ee
with normalization constants defined as 
\begin{equation}
\frac{1}{N_1}=\sum_{x} \frac{1}{x^{\alpha}}, \quad
\frac{1}{N_2}= \frac{1}{\ell-1}\sum_{x<x'} \frac{1}{(x'-x)^{\alpha}}
\end{equation} and interaction effectively being {\em short range} $N_{1,2}={\cal O}(\ell^{0})$ for $\alpha > 1$. Power-law decaying 1-spin terms and 2-spin interactions are motivated by a requirement for the spectrum of $H_0$ to be non-degenerate and free from any other discrete symmetry, which should be a generic situation.
Our results are not sensitive to the exact choice of $\alpha$, as long as we are sufficiently far away from, either strictly local interactions $\alpha=\infty$,
{or very long-ranged interactions $\alpha\approx 0$, where the model becomes mean-field like and describable by a single semi-classical degree of freedom.}

In order to avoid the need of ensemble averaging we define a time-integrated
spectral form factor as 
\begin{equation}
K_\text{int}(t) = \sum_{\tau=1}^t K(\tau),
\end{equation} 
shown in Fig.~\ref{fig:K2int}, which is indeed a self averaging quantity as demonstrated in the inset. We observe very good agreement with the RPM.
As $K_{\text{int}}(t)$ propagates deviations at short times to longer times, we also show the non-self averaging $K(t)$ at short times in Fig.~\ref{fig:K2} and again observe very good agreement with RPM upon averaging over an ensemble of values of parameter $J$ or taking a moving time-average over a short window of time for fixed $J$ (and $a,b,\alpha$). Even the fluctuations of the two averages around the theoretical prediction (RPM) for similar statistical sample size ($n=10$) look quantitatively comparable.

\begin{figure*}[ht!]
	\centering
	\includegraphics[width=0.5\textwidth]{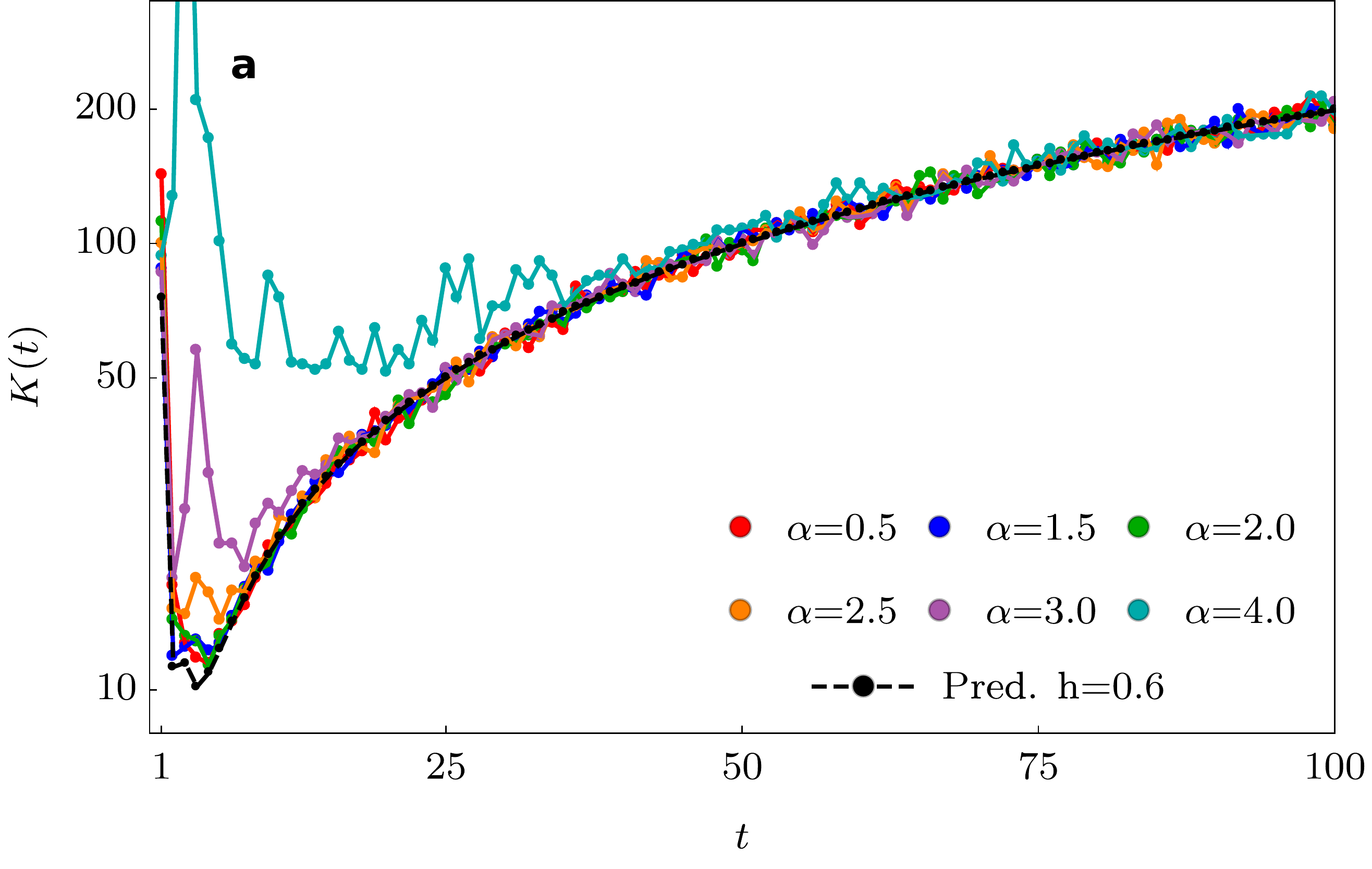}
	\includegraphics[width=0.49\textwidth]{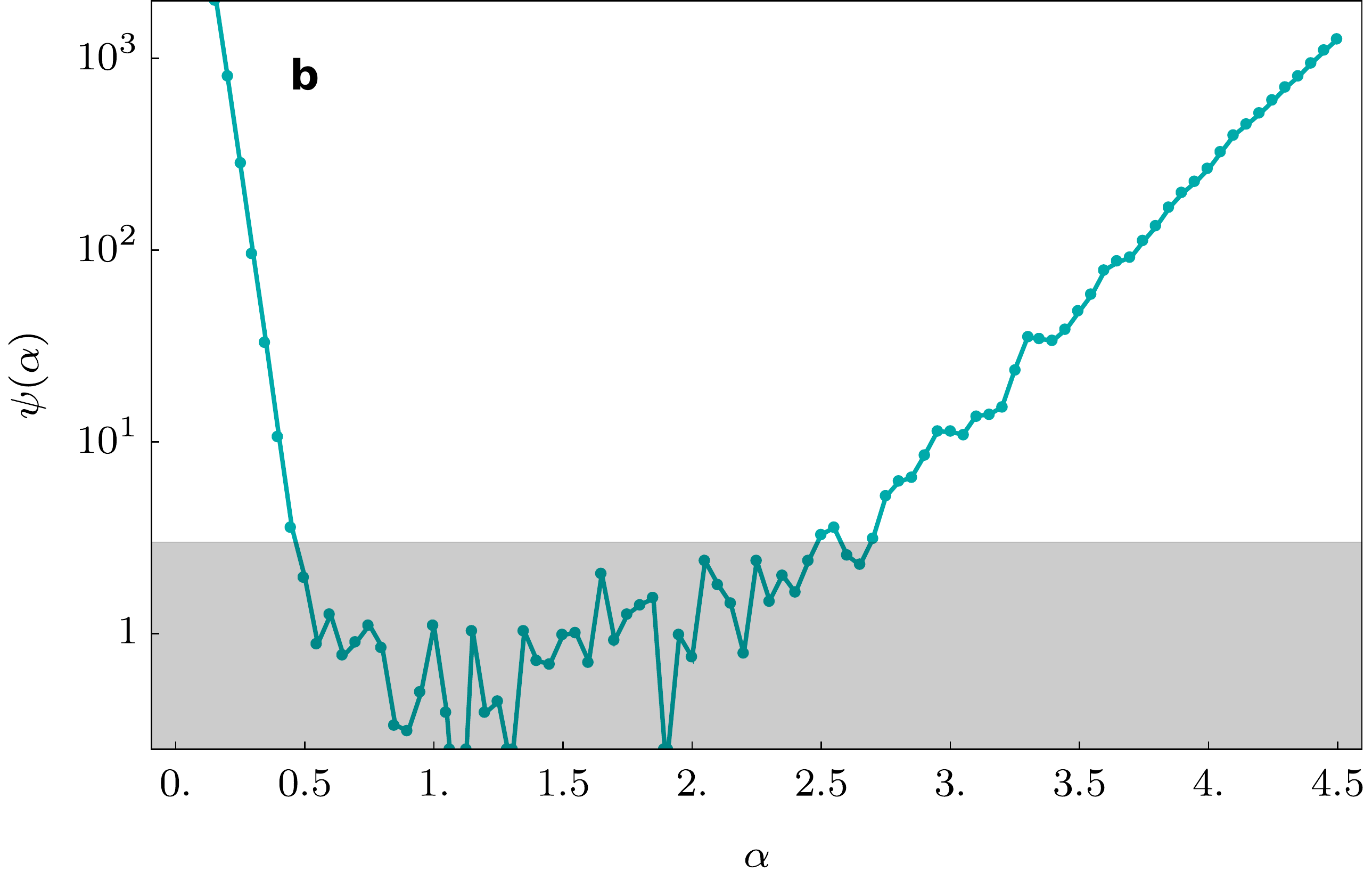}
	\caption{
{	
	Panel (a): Spectral form factor for the kicked Ising model at short times for different values of locality exponent $\alpha$. When the model becomes increasingly short ranged, it develops deviations at short times different from the prediction of the random phase model.
	Panel (b): The order parameter as defined in Eq.~(\ref{eq:order_p}) for different values of $\alpha$, which shows that the model starts to develop deviations from the random phase model when $\alpha \lesssim 0.5$ or $\alpha \gtrsim 2.5$. The shaded area guides the eye to the section with small, statistically insignificant deviations.
Averaging over 500 realizations of $J\in[5.5,255]$ is performed, other parameters are fixed to $\ell=14$, $a=1$, $b=5$ and $t_{\rm max}=100$.
\label{fig:fig4}	
}
	}
\end{figure*}

{
The data reported in Figs.~\ref{fig:K2int},\ref{fig:K2} were obtained for locality exponent $\alpha=3/2$. 
In Fig.~\ref{fig:fig4} we investigate the role of $\alpha$ in more detail. 
When the model becomes increasingly short ranged, i.e. increasing $\alpha$, the fluctuations part in Eq.~\ref{eq:delta} can no longer be neglected and the model develops deviations from RPM. 
The crossover time $t^*$ at which $K(t)$ begins to follow RMT becomes larger and starts to depend on the parameter $J$. 
But even for strictly local, nearest-neighbor interactions ($\alpha \to \infty$) at fixed $J$ this time seems to scale polynomially, perhaps like $\propto \ell^2$, 
and is still much smaller than the Heisenberg time $t_{\rm H}=2^\ell$. 

In order to quantify a transition between RPM and non-RPM physics we define the following order parameter:
\begin{align}
\psi(\alpha)&=\Big|\sum_{\tau=1}^{t_{\rm max}} \frac{K_{\rm RPM}(\tau) - K(\tau)}{K_{\rm RPM}(\tau)}\Big|.
\label{eq:order_p}
\end{align}
Since the RPM prediction for long times becomes equivalent to RMT and is expected to match $K(t)$ well, provided the model is non-integrable and ergodic,
the order parameter $\psi$ becomes independent of $t_{\rm max}$ as long as $t^* \ll t_{\rm max} \ll t_{\rm H}$. 
As shown in Fig.~~\ref{fig:fig4}, we indeed find a phase-transition-like behavior of $\psi(\alpha)$ with two, short-range and long-range critical points, $\alpha^*_{\rm long} \sim 0.5$ and $\alpha^*_{\rm short}\sim 2.5$ : For $\alpha^*_{\rm long} < \alpha < \alpha^*_{\rm short}$,
agreement with RPM is excellent and $\psi(\alpha)$ is small and of the order of expected statistical fluctuation due to averaging over $J$, 
while outside the range $\psi(\alpha)$ quickly grows.
}

\section{Conclusion}

Our work discloses the first theoretical mechanism which connects RMT to simple many-qubit systems in (effectively) low dimensions. There are many immediate
further questions and generalizations which are to be studied: i) The assumption of the pure $\delta$-correlator of phases (\ref{eq:delta}) is on the same level of rigour as the random-phase
approximation in standard semiclassics, but one may hope to find a more rigorous justification here. ii) The interesting case of local Ising interactions ($\alpha=\infty$) also obeys RMT physics \cite{Pineda} but needs to be studied separately as r.h.s. of eq. (\ref{eq:delta}) then acquires extra systematic contributions. 
iii) One may generalize our technique to study universal behaviour of dynamical correlation functions (i.e. spin structure factors, etc) in the quantum chaotic regime,
iv) Furthermore, one may expand our methods by introducing quenched disorder, say in the transverse field, and attempt to approach the many-body localization transition \cite{Huse} from the ergodic side, {for instance, by tuning the locality exponent $\alpha$.}

Our results have a direct relevance for understanding the vast body of numerical experiments, simulations, and in the near future possibly also 
experimental spectra of highly excited simple many-body systems, which correspond to ever longer accessible observation times of perfectly coherent out-of-equilibrium quantum systems
(see e.g. \cite{Bloch}). The ideas of many-body quantum chaos and random matrix theory are also vividly debated in the context of high energy physics and holography \cite{Susskind,Maldacena,Polchinski}, where our results and methods could also be applied. 

After our work has been submitted for publication, we have learned of a {series of related works} \cite{Chalker,Chalker2}, where the RMT spectral form factor has been computed for local Haar-random unitary {nearest-neighbor} quantum circuit propagator, 
{which corresponds to UE universality class of RMT}, in the limit of large local Hilbert space dimension {$q$. 
It is remarkable that in Ref.~\cite{Chalker2}, where the authors consider a related variant of RPM, but insisting on the strict locality of the (nearest-neighbour) interaction at the expense of having to consider a large $q$ limit, they found the exact same scaling of Ehrenfest/Thouless time scale $t^*\propto\log\ell$.}

	\begin{figure*}[ht!]
	\centering
	\includegraphics[width=0.98\textwidth]{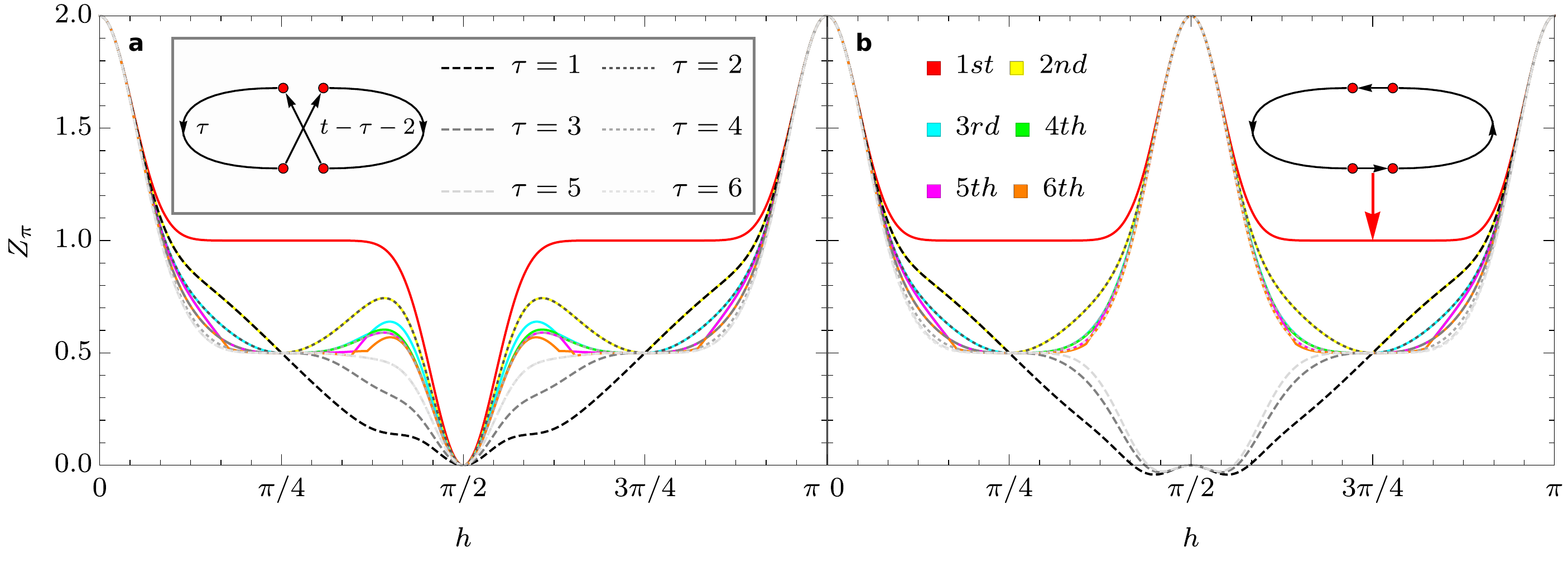}
	\caption{Partition function $Z_\pi$ of the twisted one-dimensional Ising model for different permutation families in dependence of the field parameter $h$.
	 Since in the expression the contribution occurs at a large power $\ell$, only the largest contribution matters in the first order (red), which corresponds to (anti)cyclic permutations. 
	 The first subleading corrections are given by the X-diagrams for $\tau=1,2$. 
	The coloured lines are numerical data for $t=13,14$ shown in a, b, respectively. Dashed lines are the exact expressions for $Z_{\text{X}}(\tau)$ given by eq. (\ref{eq:Xdiagram}). Note that all terms have
	multiplicities which are multiples of $2t$.
\label{fig3}	
	}
\end{figure*}

\bigskip

\noindent{\bf Acknowledgements.}	
	We thank B. Bertini, M. Horvat and B. \v Zunkovi\v c for discussions and T. H. Seligman for useful comments on the manuscript. 
	The work is supported by Advanced grant of European Research Council (ERC) 694544 -- OMNES, as well as the grants P1-0044, N1-0025 and N1-0055 of Slovenian Research Agency. 


	\section{Appendixes}

	{\bf Appendix A: Diagrammatic expansion of the leading corrections.--} 
\label{app:appC}	
When computing $K(t)$, eq.~(\ref{eq:Sum_of_permutations}), we have to sum over all permutations $\pi\in S_t$.
As we already noted, the (anti-)cyclic permutations together with the identity permutation, which form a subgroup of $S_t$, yield identical
leading contributions which become exponentially (in $\ell$) dominant in the thermodynamic limit.
Here we identify and explicitly calculate the contributions of the permutations which yield the leading (first order) corrections.
From eq.~(\ref{eq:Z_pi}) it follows that each contribution to $Z_\pi$ depends only on the number of domain walls in periodic strings $\underline{s}=(s_1,s_2, \dots,s_t)$ and $\pi(\underline{s})=(s_{\pi_1},s_{\pi_2}, \dots,s_{\pi_t})$.
Therefore $Z_\pi$ depends solely on a diagram obtained by plotting a directed graph of sequentially arranged nodes $(s_1,s_2, \dots,s_t)$ with the links
$s_{\pi_1}\to s_{\pi_2}, s_{\pi_2}\to s_{\pi_3},\ldots,s_{\pi_t} \to s_{\pi_1}$. For example, the (anti-)cyclic permutations are then represented as circular loops, meaning that they 
preserve sequential order. The next to leading order $Z_{\text{X}}$ comes from the so called X-diagrams (shown in the diagrammatic expression below and illustrated numerically in fig.~\ref{fig3}), where all connections apart from two (order changes) are kept intact (sequential). The diagrammatic expression for the case where the first sequential stretch has length $\tau$, and the second length $t-\tau-2$, reads:
\begin{widetext}
		\be		
	Z_{\text{X}} (\tau)= \!\!\!\!\!\!\sum_{s_\tau,s_{\tau+1},s_{t-1},s_t} \!\!\!\!\!\!\!\!\ T^\tau_{s_t,s_\tau}T^{t-\tau-2}_{s_{\tau+1},s_{t-1}} v_{s_\tau,s_{\tau+1}} v^*_{s_\tau,s_{t-1}} v_{s_{t-1},s_t} v^*_{s_{\tau+1},s_t}
= \frac{1}{2}\left(1+\lambda^\tau+\lambda^{t-\tau-2}-\lambda^{t-2}+\lambda^t\right).
	\label{eq:Xdiagram}
	\ee
	\end{widetext}
\phantom{??}
\vspace{-1.5cm}
\be
\hspace{2.5cm}\vcenter{\includegraphics[width=3cm,trim={0.8cm 0.1cm 0.8cm 0.1cm},clip]{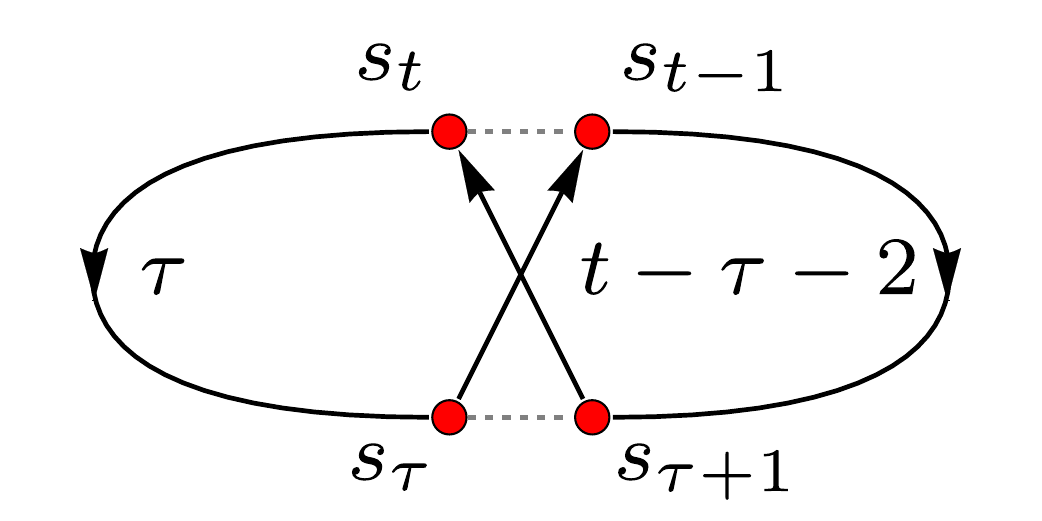}}\nonumber
\ee
	Circular black arcs represent summations over stretches of sequential spins $[\tau',\tau'']$, namely over all $s_{\tau},\tau' <\tau<\tau''$. These are given by the powers of the transfer matrix
        $T^{\tau''-\tau'+1}_{s_{\tau'},s_{\tau''}}$, specifically $T_{ss'}^p = \frac{1}{2}\left(1+ (-1)^{s-s'}\lambda^p \right)$ where $\lambda=\cos 2h$ plays the role of the coupling constant.         
        Red dots correspond to remaining spins which one still needs to sum over while putting matrix element $v_{ss'}$ for each broken sequential link $s\to s'$ (dotted) and $v^*_{ss'}$ for each crossed link $s\to s'$.

Of similar importance are the XX-diagrams with two sequential crossings:
\be\centering\vcenter{\includegraphics[width=3cm,trim={0.8cm 0.1cm 0.8cm 0.1cm},clip]{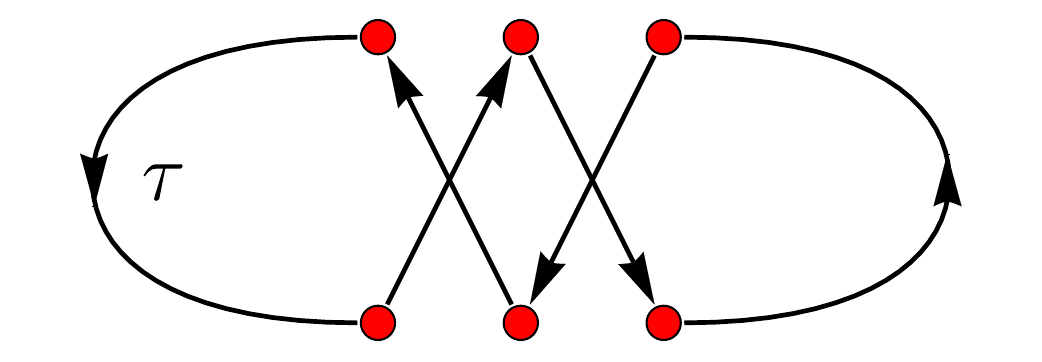}}\nonumber\ee
which are straightforwardly evaluated
	\be
	Z_{\text{XX}}(\tau)=\frac{1}{2}\left(1+\lambda^{\tau+2}+\lambda^{t-\tau-2}+2\lambda^{t-4}(\lambda^4\!-\!\lambda^2\!+\!\frac{1}{2})\right).
	\ee
One can show that contributions of all other diagrams, starting with two crossings separated by sequential stretches, triple crossings, etc., are of the form $Z_{\text{other}}(\tau) = 2^{-n}(1+ {\cal O}(\lambda^k))$
with $n\ge 2$ and $k\ge 1$, so they contribute to $Z^\ell$ only beyond the second order in $t/2^\ell$ and will be ignored here.

Each X and XX diagram (for fixed $\tau$) has multiplicity $t^2$ (or $t^2/2$ for $\tau=t/2$), since any $Z_\pi$ is invariant under $2t$ cyclic and anti-cyclic permutations and we can start drawing the diagram at 
$t/2$ inequivalent points. 

We now see that the smallest gap of $Z_{\text{Id}}-Z_\pi > 0$ comes from $Z_{\text{X}}(1)$ for $\lambda>0$ and $Z_{\text{X}}(2)$ for $\lambda<0$ (see fig.~\ref{fig3}).
Since these terms enter to power $\ell$, the sub-leading contributions to $K(t)$ are exponentially suppressed by a factor of the order of 
$\sim t ((1+\lambda)/2)^\ell$  for $\lambda>0$ and $t ((1+\lambda^2)/2)^\ell$ for $\lambda<0$. 
When we approach the Heisenberg time $t\sim t_{\rm H}=2^\ell$ these contributions become important, as we will see in the next section.

{\bf Appendix B: Second order term in $t/t_{\rm H}$.--}	
	RMT predicts that the next term in the expansion is $-2t^2/2^\ell$, eq. (\ref{RMTSFF}). We now show that the contributions of the X and XX diagrams and the possible repetitions of the spin configurations almost cancel, yielding exactly the RMT result.
	
	In the first order approximation in $t/2^\ell$, we neglected the possibility that the spin configurations $\ul{s}_\tau$, $\tau=1,\ldots,t$ can repeat after some time. The leading order correction to this consist from cases with a single repetition $\underline{s}_{\tau_1}=\underline{s}_{\tau_2}$. Then, eq. (\ref{eq:delta}) renders the permutation $\pi$ to run over a factor group $S_t/S_{\{\tau_1,\tau_2\}}$, where $S_{\{\tau_1,\tau_2\}}$ is a two-element permutation group. Because our leading order sum (\ref{eq:Sum_of_permutations}) still runs over the entire permutation group $S_t$ we end up counting each element twice, so we need to subtract the over-counted terms
	\bea
	K(t)&=&\sum_{\pi\in S_t} Z^\ell_\pi - \sum_{1\le \tau_1<\tau_2\le t} \frac{1}{2} \sum_{\pi\in S_t} \sum_{\underline{s}_1, \dots, \underline{s}_t}^{\underline{s}_{\tau_1}=\underline{s}_{\tau_2}} F_\pi(\ul{\ul{s}})\nonumber
	\\
	&=&\sum_{\pi\in S_t} Z^\ell_\pi-
	\frac{1}{2} \frac{t}{2} \sum_{\tau=1}^{t-1} \sum_{\pi \in S_t} \sum_{\ul{\ul{s}}}^{\underline{s}_\tau=\underline{s}_{t}} F_\pi(\ul{\ul{s}}),
	\eea 
	where $F_\pi(\ul{\ul{s}})= \prod_{x=1}^\ell  \prod_{\tau=1}^t v_{s_{x,\tau},s_{x,\tau+1}}v^*_{s'_{x,\tau},s'_{x,\tau+1}}$ and $\ul{\ul{s}}\equiv (s_{x,\tau};1\le x\le\ell,1\le \tau\le t)$.
	In the second line we use time invariance to set $\tau_2=t$ and $\tau_1=\tau$. 
	Since the repeated spin contributes the same regardless of its value, the sum trivializes for $\pi=\text{Id}$ to a product of sums for separate spins, each contributing 
	\be
	\sum_{\ul{s}}^{s_\tau=s_{t}} F^{\ell=1}_\pi(\ul{s})=2T_{00}^\tau T_{00}^{t-\tau}
	=\frac{1}{2}(1+\lambda^\tau+\lambda^{t-\tau}+\lambda^t).
	\ee	
	Following the same argument as before, invariance under cyclic and anti-cyclic permutations again yields multiplicity $2t$.
         But because of the repetition of spin configurations, the X and XX diagrams where the crossed link contains the repeated spin yield the contribution in the same order $\frac{1}{2}(1+{\cal O}(\lambda^\tau))$: 
	\begin{widetext}
	\begin{align}
	K(t)&=2t(1+\lambda^t)^\ell+ \frac{t^2}{2^{\ell}} 
	\Bigg(\sum_{\tau=1}^{t-3}(1+\lambda^\tau+\lambda^{t-\tau-2}-\lambda^{t-2}+\lambda^t)^\ell
	&	\vcenter{\includegraphics[width=2.4cm,trim={0.8cm 0.1cm 0.8cm 0.1cm},clip]{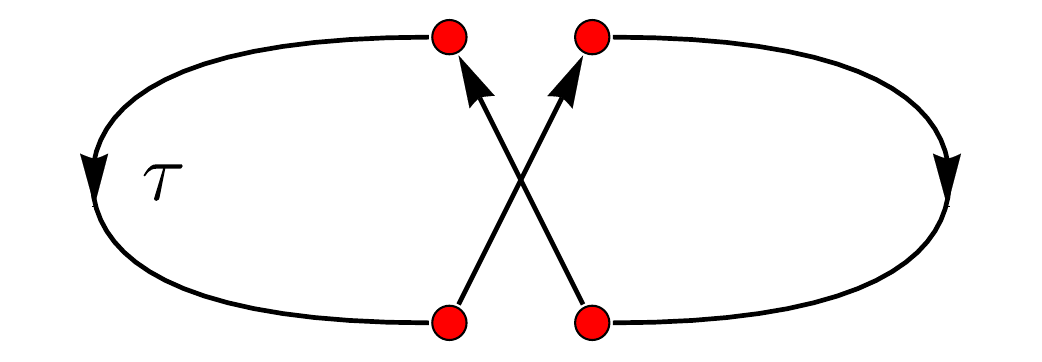}}\hspace{-15.6cm}		
	\nonumber\\
	&+
	\sum_{\tau=3}^{t-3}(1+\lambda^{\tau}+\lambda^{t-\tau}+\lambda^{t-4}(2\lambda^4-2\lambda^2+1))^\ell
		&
		\vcenter{\includegraphics[width=2.4cm,trim={0.8cm 0.1cm 0.8cm 0.1cm},clip]{diagram2.pdf}}\hspace{-15.6cm}
	\nonumber\\
	&-
	\sum_{\tau=1}^{t-1} (1+\lambda^\tau+\lambda^{t-\tau}+\lambda^t)^\ell 
	&	\vcenter{\includegraphics[width=2.4cm,trim={0.8cm 0.1cm 0.8cm 0.1cm},clip]{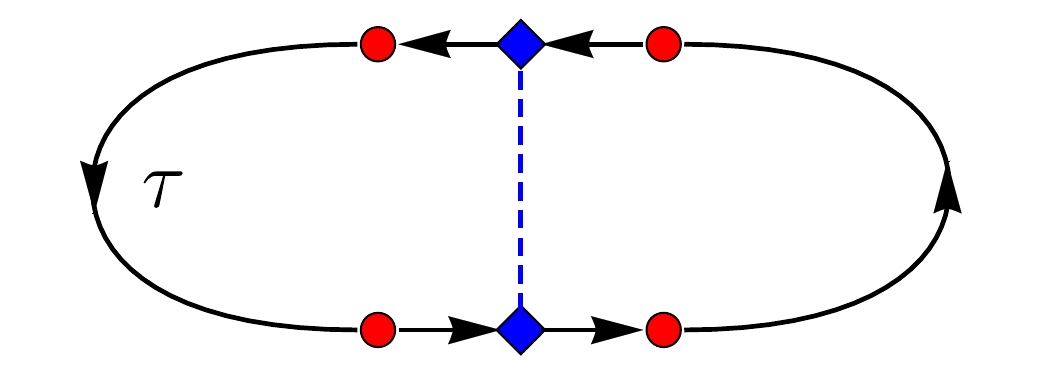}}\hspace{-15.3cm}	
	\vcenter{\includegraphics[width=2.4cm,trim={0.8cm 0.1cm 0.8cm 0.1cm},clip]{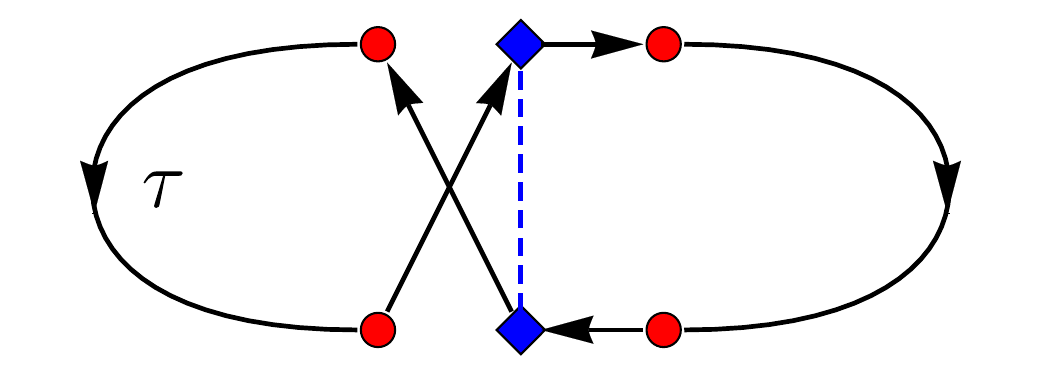}}\hspace{-15.6cm}
	\nonumber\\
	&-
	\sum_{\tau=3}^{t-3} (1+\lambda^\tau+\lambda^{t-\tau}+\lambda^t)^\ell
\Bigg)
	+... \nonumber 
	&	\vcenter{\includegraphics[width=2.4cm,trim={0.8cm 0.1cm 0.8cm 0.1cm},clip]{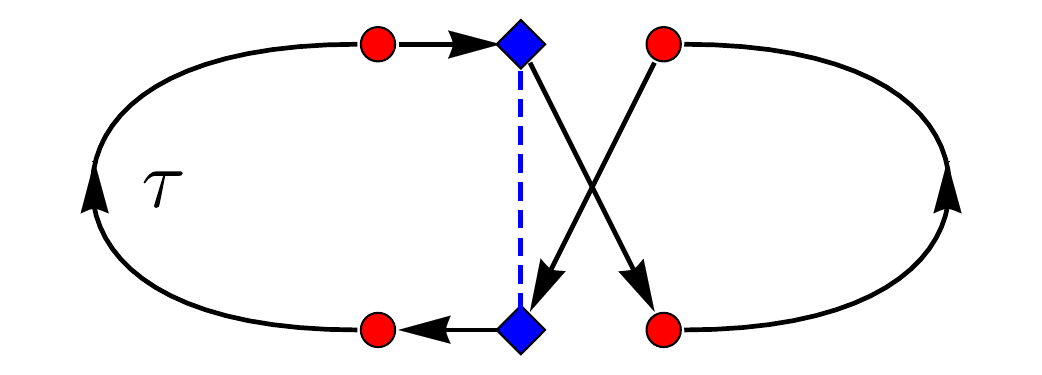}}\hspace{-15.3cm}	
	\vcenter{\includegraphics[width=2.4cm,trim={0.8cm 0.1cm 0.8cm 0.1cm},clip]{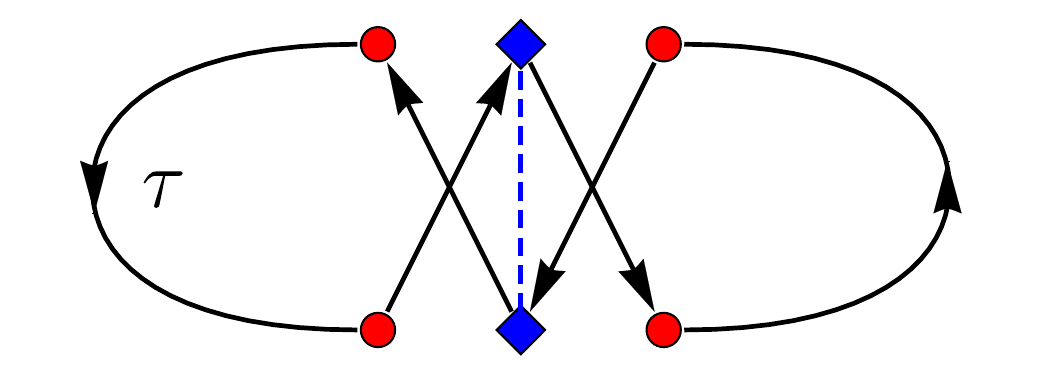}}\hspace{-15.6cm}	
	\nonumber\\
	&= 2t- \frac{2 t^2}{2^{\ell}} + {\cal O}\left(\frac{t^3}{4^\ell}\right). \label{eq:canc}
	\end{align}
	\end{widetext}
On the right, the corresponding diagrams are shown for clarity, where the blue diamond sites connected with a dashed line depict the repeating spin. The first two sums come from enumerating all X and XX diagrams (as explained in Appendix C). The sum from the repeating spin configurations comes next, and is written in two parts.
 The combinatorial factor of the diagrams is $4t$ when $\tau=1,2,t-1,t-2$ and $8t$ otherwise, which we take into account by writing two sums with different starting and final value of $\tau$. In the last line of
 eq. (\ref{eq:canc}) we note a remarkable cancellation of all terms apart from the RMT result for times $t > t^\diamond = {\cal O}(\ell \log\lambda)$ where $t_{\rm H}/t^\diamond$ is still exponentially large in $\ell$.
 This could be viewed as a quantum many-body analogy of the Sieber-Richter self-encountering-orbits mechanism.



\end{document}